\documentstyle[12pt]{article}
\begin{document}
\begin{titlepage}

\begin{centering}

{\LARGE On the geometry and conformation of starburst dendrimers}\\
\vspace{1cm}
{\large Antonio de P\'adua}\\
Departamento de F\'{\i}sica, UFPE\\
50670-901 Recife, PE, Brazil\\
\vspace{0.5cm}
{\large J. A. de Miranda-Neto}\\
Department of Physics, Carnegie Mellon University\\
Pittsburgh, PA 15213\\
\vspace{0.5cm}
{\large Isaac M. Xavier Jr.$^{\diamond}$}\\
Department of Chemistry, University of Pennsylvania\\
Philadelphia, PA 19104-6323\\
\vspace{0.5cm}
{\large  and Fernando Moraes$^{\star}$}\\
School of Natural Sciences, Institute for Advanced Study\\
Princeton, NJ 08540\\

\end{centering}
\begin{abstract}

In this work, we propose a geometrical model for the study of conformational
properties of a starburst dendrimer with the topology of a truncated Bethe
lattice. A convenient embedding of the Bethe lattice in the hyperbolic plane is
used to study the architecture of the dendrimer. As results, we find an upper
bound for the molecular size and the density profile.

\end{abstract}

\noindent
\\
$^{\diamond}$Permanent address:\hspace{5cm}${\star}$Permanent address:\\
Departamento de Qu\'{\i}mica Fundamental, UFPE\hspace{0.5cm}Departamento de
F\'{\i}sica,UFPE\\
50670-901 Recife, PE, Brazil\hspace{4cm}50670-901 Recife, PE, Brazil \\


\end{titlepage}
\def\carre{\vbox{\hrule\hbox{\vrule\kern 3pt
\vbox{\kern 3pt\kern 3pt}\kern 3pt\vrule}\hrule}}

\baselineskip = 18pt

\section {Introduction}

Recently, highly branched molecules called dendrimers or arborols have become
increasingly important in macromolecular science [1-8]. Dendritic
macromolecules are three-dimensional polymers that emanate from a central core,
have a
defined number of generations and functional end groups, and are synthesized
in a stepwise way by a repetitive reaction sequence [1]. An important
difference between linear polymers and dendrimers is that a linear polymer
consists of an entanglement of single molecular chains. In a dendrimer, by
contrast, the many branches give rise to a very high number of terminal
functional groups in each molecule. In spite of their complexity, it is now
possible to devise synthetic strategies which enable a systematic molecular
morphogenesis from small molecules to macroscopic matter with control of size,
shape, topology, flexibility and surface chemistry [1]. New synthetic paths
can lead to materials with promising properties and potential new applications
for catalysis, selective binding, electron transfer and energy conversion.

Dendrimers prepared from rigid units have a more precisely defined three-
dimensional structure compared to their flexible counterparts. The rigid
repeated units impose geometrical restrictions and thus little conformational
freedom. Such freedom comes mainly from the rotational angles which depend on
steric effects and/or hydrogen bonding [1]. The branching angle, on the other
hand, is determined by the chemical nature of the branching juncture. In spite
of the steric strain be quite dramatic in these dendritic systems [9], a new
all-hydrocarbon rigid macromolecule, $C_{1134}H_{1146}$, has been recently
synthesized and characterized [10]. This dendrimer consists of 94
phenylacetylene monomer units (94-mer in polymer speech). Its
two-dimensional representation displays a snowflake-like skeleton (as shown in
figure 1).

This 94-mer starburst dendrimer presents some outstanding geometrical
properties. Its two-dimensional skeleton is isomorphous to the
three-coordinated Bethe lattice [11].  Also, computer generated
three-dimensional models exhibit a globular shaped surface with large voids and
niches in its interior [10] reminiscent of a negatively curved surface. As we
shall see, this implies that not only the topology of these molecules is that
of the Bethe lattice, but also their  geometry and symmetries. The conventional
Bethe lattice is an infinite ramified network without rings and unable to fit
entirely in the
three-dimensional space since its number of vertices grows exponentially with
the distance, while the available space grows only as a power law. Usually,
it is viewed as a lattice in an infinite-dimensional Euclidean space. However,
the Bethe lattice can be embedded in a constant negative-curvature surface, the
Lobachevsky or hyperbolic plane [12-14]. There, it becomes a regular lattice
with well defined bond angles and distances. Although a theorem due to Hilbert
[15] forbids the entire hyperbolic plane to be isometrically immersed in the
three-dimensional Euclidean space, a result of Cartan [16] permits a finite
portion of the hyperbolic plane to do so. This appears as a wrinkled surface:
the larger is the portion, the more crumpled it looks like. The maximum size
would be that one which beyond it the immersed surface begins to
self-intersect, producing the geometrical analogue of steric strain.

In this article, we propose a hyperbolic dendritic model for the 94-mer
starburst dendrimer in particular, but some of the results are general enough
to apply to other rigid systems. The model enables us to study structural
properties such as the dendrimer scaling behavior, its intramolecular density
profile, and gives an upper bound for its molecular size.

\section {Hyperbolic Model}

The Bethe lattice has been widely used as a
theoretical substrate for the modeling of many physical systems. It is defined
as an infinite ramified set of points, each one connected to q neighbors,
such that no closed rings exist, forming a completely open hierarchical
structure. The natural hierarchy presented in such lattice provides the
analytical treatment of a broad spectrum of theoretical problems, many of them
analytically unsolvable when modeled in ordinary Euclidean lattices. On the
other hand, behind those simple analytical solutions hide unusual geometrical
characteristics [17]. By construction, the total number of vertices in the
Bethe lattice grows exponentially with the distance. In this way, this lattice
can be globally realized in an Euclidean space only if the space is of infinite
dimension.

Perhaps a a little explanation is due here. One might try to build the lattice
according to the following steps: i) take a starting point on a given plane;
ii) draw three straight line segments of a given length from that point such
that the angle between each neighboring pair is always 120 degrees; iii) at the
end of each segment repeat step (ii) except that one needs not to redraw
segments already existing; iv) go back to (iii). It is fairly obvious that this
leads to the honeycomb or hexagonal lattice (the crystalline structure of
graphite). If one would like to avoid the rings, while keeping constant bond
length and angles, very soon one would need to step out of the plane and keep
building the lattice in three-dimensional space. Soon enough, the same kind of
problem would be happening in three-dimensional space and one would need to
step into the fourth dimension. And so on. Then, in order to build a Bethe
lattice with an infinite number of of vertices, with equal bond !
 lengths and angles, one would need an infinite-dimensional Euclidean space.

It is clear then, that when one tries to build the Bethe lattice on a flat
surface, the result is something altogether different (honeycomb) or one would
have to keep increasing the dimensionality of space while the lattice grows.
Another alternative was pointed out by Mosseri and Sadoc [18]: if you allow for
the surface to be curved in a certain way, then the lattice may be built to any
desired extension. This curved surface has constant negative curvature and is
known in the mathematical literature as the two-dimensional hyperbolic space or
hyperbolic plane [19]. Whereas a surface of constant positive curvature is a
sphere, the corresponding constant negative curvature surface locally looks
like a saddle. While the sphere has a finite area, the hyperbolic plane is
infinite. Built on the surface of the hyperbolic plane, the Bethe lattice can
be
seen as a regular crystallographic lattice, presenting well defined angles
between bonds and with spatial coordinates assigned to its vertices. In the
hyperbolic plane, such lattice is just a regular tiling of the ``hyperbolic
floor" by polygons of an infinite number of edges [18] just like the honeycomb
is a tiling of the plane by hexagons. The details on the geometry and
symmetries of the hyperbolic Bethe lattice may be found in references [12,13].

A hierarchical algorithm for constructing the Bethe lattice on the hyperbolic
plane, yielding the coordinates of the vertices, has been described in detail
in reference [13]. The computation of the distance between, say a given vertex
and the origin involves geometrical concepts that do not properly fit in here,
the interested reader may find the details in [12]. A word of caution: in
statistical mechanics one usually deals with the topological distance, which is
the  number of bonds between two given sites of the Bethe lattice. This is {\em
not} what we do here. We use a physically more realistic distance: the actual
geometrical distance as defined in hyperbolic space.

With the knowledge of spatial coordinates plus the possibility of defining
geometrical distances between sites, one is able to analyze the scaling
properties
involving the way the number of volume ($N$) and surface ($n$) sites grow when
the branched structure increases its size. For the purely topological
q-coordinated Bethe lattice, it is well known that the total number of lattice
points within a
distance of $t$ steps from the origin is
 \begin{equation}
N=[q(q-1)^t-2]/(q-2)
\end{equation}
and the number of points at the lattice surface for the same distance $t$ is
\begin{equation}
n=q(q-1)^{t-1}.
\end{equation}
Therefore, for large $t$, the fraction $n/N$ (surface/volume) is:
\begin{equation}
 {n\over N}={q-2 \over{q-1}}.
\end{equation}

Thus, for a conventional Bethe lattice of any coordination number, the fraction
$n/N$ tends to a finite limit. This feature is quite different for finite
dimensional Euclidean lattices, where the fraction $n/N$ goes to zero in the
thermodynamic limit. In the hyperbolic case we recover [17] the usual zero
limit for n/N at large distances. This is shown in figure 2.

\section {Results and Discussion}
In this paper, the phenyacetylene dendrimer architecture is modeled by using a
hyperbolic Bethe lattice that shares structural similarities to its
two-dimensional structural formula. In this way, each one of the 94 benzene
nuclei corresponds to a point in the three-coordinated Bethe lattice which is
truncated after its fifth propagation. The nuclei are connected by rigid
acetylene units. The 48 peripheral benzene nuclei are replaced by
3,5-di-tert-butylphenyl groups.

We base our model on the assumption that the equilibrium configuration of the
dendrimer is that of minimum energy. Considering that the building blocks of
the dendrimer are two-dimensional objects and that steric hindrance promotes an
increase of the energy, the lowest energy requirement equals a maximum spread
under the restriction of fixed branching angle and spacer length. In other
words, the two-dimensional surface\footnote{Not to be confused with the
dendrimer outer surface.} defined by the dendrimer must be of maximal area and
at the same time must permit fixed branching angles and spacer length. As seen
in the previous section, the hyperbolic plane permits the growth of a structure
isomorphous to the dendrimer and also with fixed bond angles and lengths, the
Bethe lattice. Moreover, the area of a circle of radius r in the hyperbolic
plane of Gaussian curvature  $ K=- \kappa ^{2}$ is given by
\begin{displaymath}
A=4\pi sinh^{2} (\kappa r/2)/\kappa^{2},
\end{displaymath}
which is much larger than the area of an Euclidean circle of same radius, $\pi
r^{2}$.  That is, our model can be viewed as  ``variational" in the sense that
the parameter $\kappa$ is tuned to accommodate at the same time an area larger
than the Euclidean one ($\kappa \not= 0$) and a commensurate three-coordinated
Bethe lattice  (from $ \kappa d = ln3$, see below).  In very simple words: if
the dendrimer, being essentially a two-dimensional molecule, due to obvious
crowding reasons cannot grow in the Euclidean plane then it chooses a
``roomier" two-dimensional space to do so.

The implication of figure 2 to our model is now evident: as the dendrimer
grows, the distribution of benzene sites (or vertices of the Bethe lattice)
over the volume of the molecule becomes much larger than the distribution on
the external surface. This is a consequence of the crumpled nature of the
Euclidean immersion of the hyperbolic plane and means that backfolding is taken
into account in our model. In fact, this is obvious if one looks at Table I of
reference [12]. The immediate consequence for the phenylacetylene dendrimer is
that 42  out of the 48,  3,5-di-tert-butylphenyl groups of the fifth
generation, are on the surface while  6 of them lie within the spread of the
previous generation.

In hyperbolic space, as on the surface of a sphere, the length scale is coupled
to the radius of curvature, $R = 1/K$, of the surface. So, for example, if one
wants to circumscribe a cube of edge length $l$, the radius of the
circumscribing sphere has to be $l \sqrt{3}/2$ and one says that the cube and
sphere are commensurate with each other. In our case, the Gaussian curvature of
the hyperbolic plane commensurate with the three-coordinated Bethe lattice of
edge (in this case the distance between two adjacent vertices) $d$, is given by
the relation $ \kappa d = ln3$ from reference [20]. Now, d is the length of the
phenylacetylene unit which is estimated to be 7.0 \AA.  With this value, the
constant $\kappa$ is 0.157 \AA$^{-1}$ and the curvature K is -0.025 \AA$^{-2}$.
The hyperbolic model also provides an upper bound for the molecular size by
considering the phenylacetylene dendrimer isomorphous to the fifth propagation
of a negatively-curved three-coordinated Bethe lattice.!
  Figure 3 depicts the total number of vertices N within a radius r from the
core, as a function of r, for the hyperbolic Bethe lattice [17]. For the 94-mer
phenylacetylene dendrimer (N=94), one obtains from that plot that r equals
4.60 (in units of $\kappa^{-1}$) which gives an upper bound radius of 29.3 \AA
\, for the molecule. This is in good agreement with the value 27.5 \AA \,
estimated from space-filling molecular models [10].

It is expected that, as the volume of a dendrimeric molecule increases
cubically,  its mass increases exponentially [1,2]. As shown in figure 3, the
total number of vertices (or molecular mass in suitable units) in our model
also increases exponentially with its metric distance to the core. Figure 4
shows the density profile obtained from figure 3 by dividing $N(r)$ by the
volume of a sphere of radius $r$. It shows a initial decrease, goes through a
minimum and from then on keeps increasing with the radius.

Differently from molecular models [21], that are able to predict the molecular
weight threshold that corresponds to the transition from an extended to a
globular shape, our model concentrates in the globular shape side of the
transition. Moreover, in order to have its maximum conformational freedom, a
necessary condition to reach the minimum energy, the dendrimer is assumed to be
in solution. The conformation of dendritic macromolecules in the solid state
and in solution is still controversial [2].  Lescanec and Muthukumar [6]
developed a model which predicts  inward folding of branch units, a density
profile maximum at the core of the dendrimers, and a distribution of terminal
groups throughout the structure. This model simulates kinetic growth of
starburst dendrimer by means of a self-avoiding walk algorithm. Klushin and
Mansfield [7] developed versions of the self-avoiding model to study polyether
dendrimers in equilibrium through Monte Carlo simulations. Their model pres!
 ents a dispersion of end groups throughout the molecule, even near the core,
something that substantially lowers the density even considering backfolding.
As a result, their density profiles decrease with the size of the molecule. De
Gennes and Hevert [22] developed a self-consistent field model which predicts a
dendrimer structure emanating radially outward with a density profile minimum
at the center of the dendrimer. In their model, all the terminal groups are on
the periphery of the macromolecule and the ideal dendritic growth will only
occur until a certain generation is reached at which point steric congestion
will prevent further growth. Molecular modeling performed by Goddard et al.
[1,21] predicts internal cavities, most of the terminal groups on the surface
of large dendrimers and an increasing density profile.  The density profile
(figure 4) obtained in this letter, quite surprisingly, shows characteristics
of all the above models! Initially it decreases with th!
 e radius, then it picks up and keeps increasing  with the radi!
 al dista

nce from the core. This is certainly a good indication of the important role of
geometrical aspects in dendrimer modeling. Of course it would be a very
na\"{\i}ve assumption to think that geometry alone solves the problem. On the
contrary, we believe our contribution indicates that there is something in
between molecular modeling/self-consistent modeling and kinetic/Monte Carlo
models and perhaps the liaison is geometry.

In this work, a hyperbolic-geometrical model has been used to investigate
starburst dendrimer architecture. Although we have focused our work on the
phenylacetylene dendrimer, most of the results are general enough to be applied
to other rigid dendrimers.

\noindent {\bf Acknowledgments}

This work was partially supported by CNPq and FINEP. IMXJ thanks the
International Center of Condensed Matter Physics, Brasilia, Brazil. Part of
this work was developed during his stay there.

\bigskip
\noindent
{\large \bf References}

\bigskip
\noindent
\begin{enumerate}
\item D. A. Tomalia, A. M. Naylor  and W. A. Goddard III, Angew. Chem. Int. Ed.
Engl. {\bf 29}, 138 (1990).
\item J. M. Fre\'chet, Science {\bf 263}, 1710 (1994).
\item J. F. G. A. Jansen, E. M. M. de Brabander-van der Berg and E. W. Meijer,
Science {\bf 266}, 1226 (1994).
\item S. M. Risser, D. N. Beratan and J. N. Onuchic, J. Phys. Chem. {\bf 97},
4523 (1993).
\item P. Biswas and B.J. Cherayil, J. Chem. Phys. {\bf 100}, 3201 (1994).
\item R. L. Lescanec and M. Muthukumar, Macromolecules {\bf 23}, 2280 (1990).
\item M. L. Mansfield and L. I. Klushin, Macromolecules {\bf 26}, 4262 (1993).
\item Y. Tamori, Phys. Rev. E {\bf 48}, 3124 (1993).
\item J. S. Moore and Z. Xu, Macromolecules {\bf24}, 5893 (1991).
\item Z. Xu and J. S. Moore, Angew. Chem. Int. Ed. Engl. {\bf 32}, 246 (1993).
\item M. Kurota, R. Kikuchi and T. Watari, J. Chem. Phys. {\bf 21}, 434 (1953).
\item J. A. de Miranda-Neto and F. Moraes, J. Phys. I (France) {\bf 2}, 1657
(1992).
\item J. A. de Miranda-Neto and F. Moraes, J. Phys. I (France) {\bf 3}, 29
(1993).
\item B. S\"{o}derberg, Phys. Rev. E {\bf 47}, 4582 (1993).
\item D. Hilbert, Trans. Amer. Math. Soc. {\bf2}, 87 (1901).
\item E. Cartan, Bull. Soc. Math. France {\bf 47}, 125 (1919).
\item A. de P\'adua, J.A. de Miranda-Neto and F. Moraes, Mod. Phys. Lett. B
{\bf 8}, 909 (1994).
\item R. Mosseri and J. F. Sadoc, J. Phys. (Paris) Lett. {\bf 43}, L-295
(1982).
\item M. J. Greenberg, Euclidean and Non-Euclidean Geometries (Freeman, San
Francisco, 1980).
\item J. A. de Miranda-Neto and F. Moraes, J. Phys. I (France) {\bf 3}, 1119
(1993).
\item A. M. Naylor and W. A. Goddard III, J. Am. Chem. Soc. {\bf 111}, 2339
(1989).
\item P. G. de Gennes and H. Hevert, J. Phys. Lett. {\bf 44}, 351 (1983).
\end{enumerate}

\newpage
\noindent
{\bf Figure Captions}
\bigskip
\noindent
\begin{enumerate}

\item The core of the 94-mer phenylacetylene dendrimer.
\item Rate between the number of surface vertices to the number of volume
vertices $n/N$ as function of the radial distance r.
\item Total number of vertices N within radius r, as a function of r.
\item Density profile for the hyperbolic model.
\end{enumerate}
\end {document}
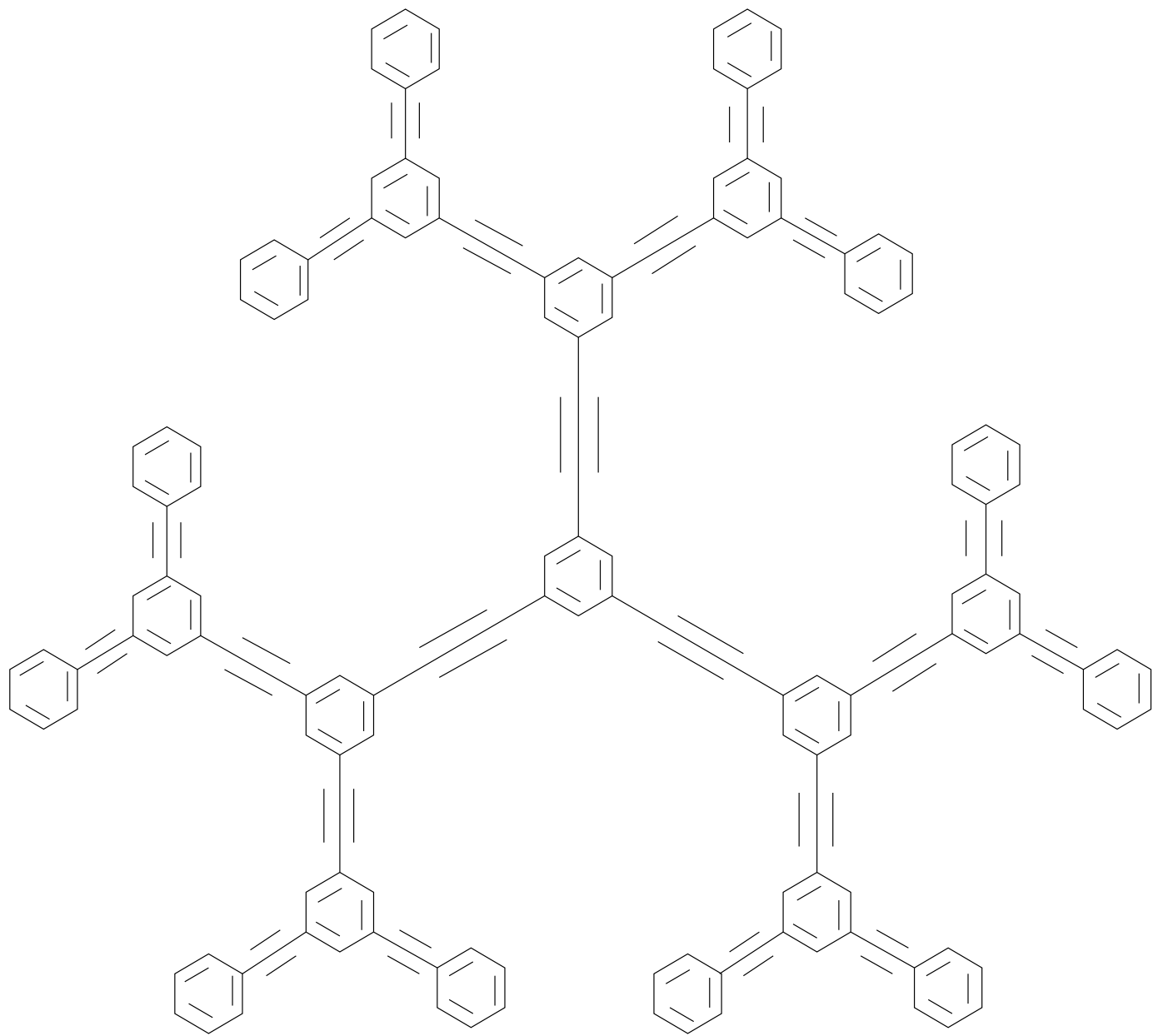
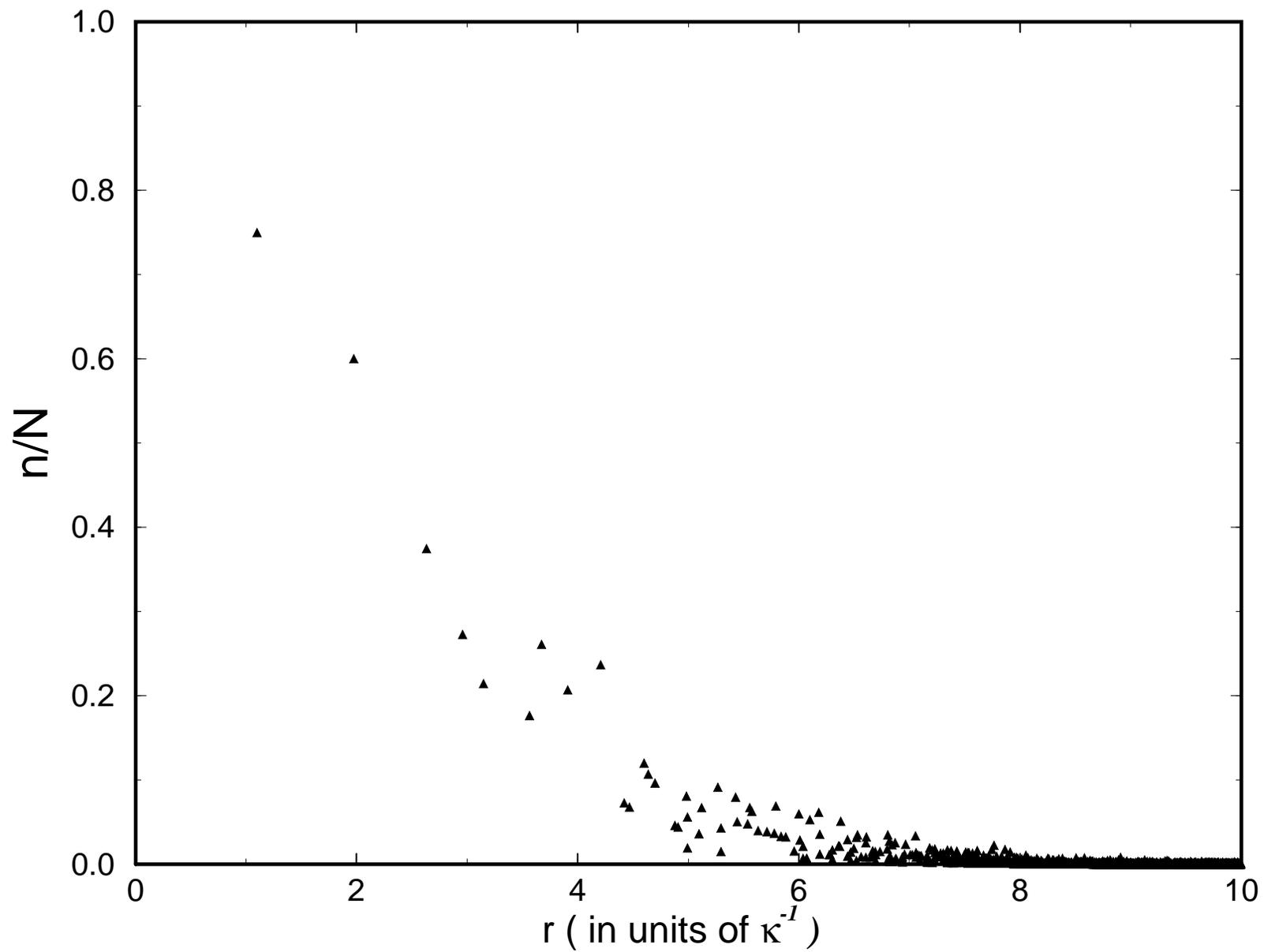
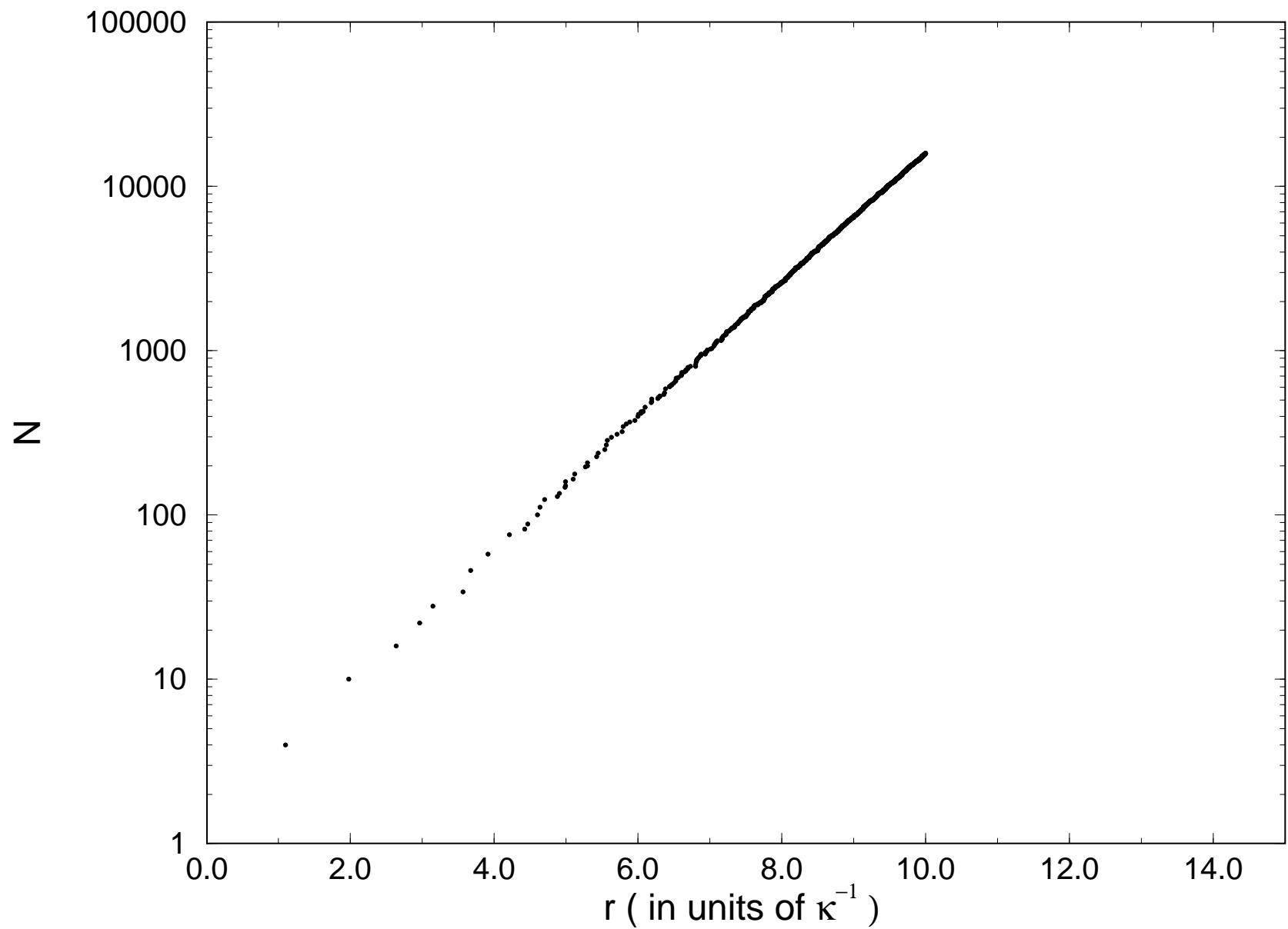
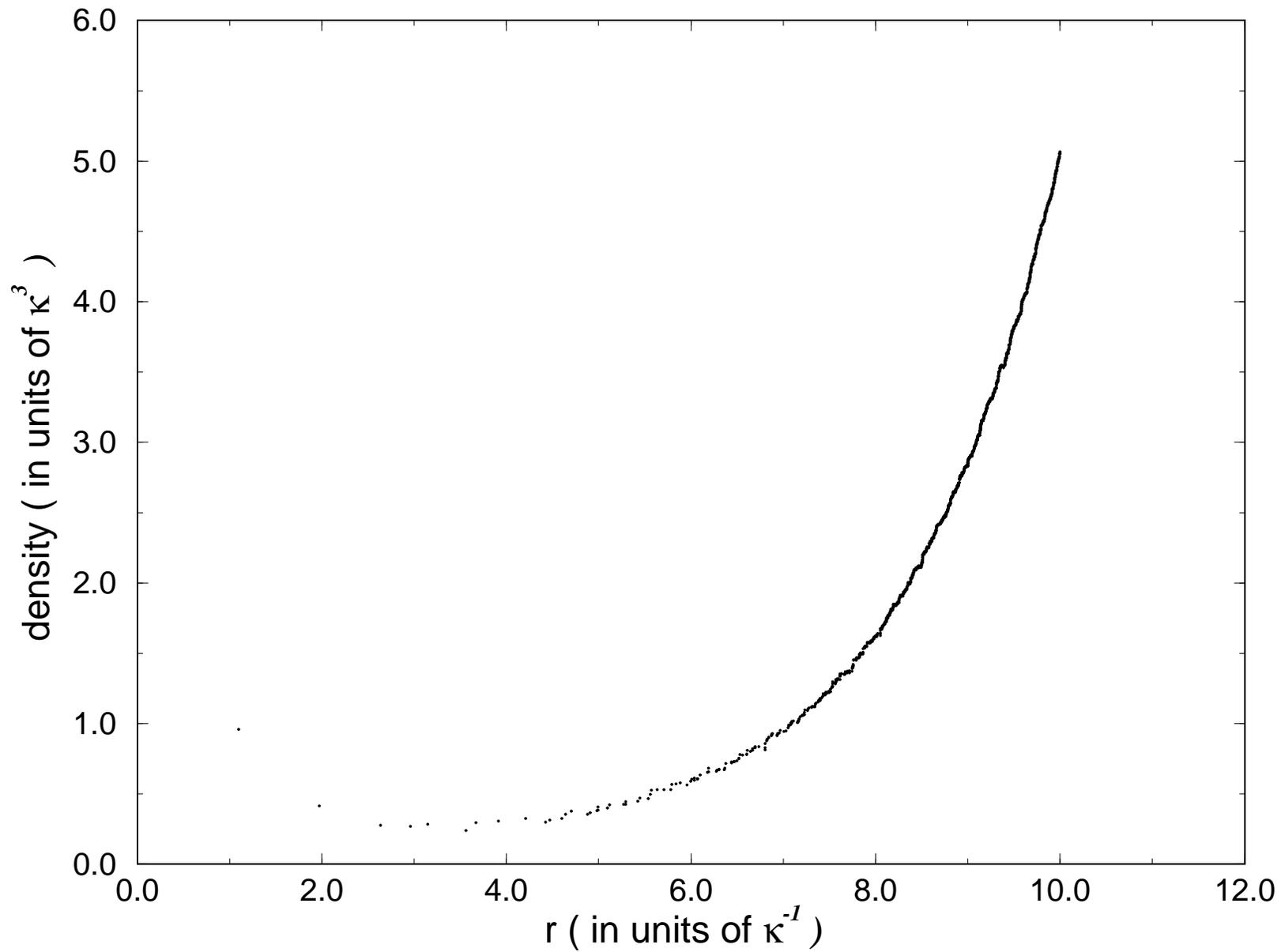
